\documentclass[floatfix,aps,prb,twocolumn,superscriptaddress]{revtex4-2} 

\usepackage[T1]{fontenc}
\usepackage{lipsum}  
\usepackage{graphicx}
\usepackage[english]{babel}
\usepackage{upgreek}
\usepackage{amsmath}
\usepackage{tikz}
\usepackage{amssymb}
\usepackage{tabls}
\usepackage{dsfont}
\usepackage[colorlinks=true,citecolor=blue,linkcolor=red,urlcolor=blue]{hyperref}
\usepackage{physics}
\usetikzlibrary{tikzmark,fit}

\makeatletter
\def\bbl@set@language#1{%
	\edef\languagename{%
		\ifnum\escapechar=\expandafter`\string#1\@empty
		\else\string#1\@empty\fi}%
	\@ifundefined{babel@language@alias@\languagename}{}{%
		\edef\languagename{\@nameuse{babel@language@alias@\languagename}}%
	}%
	\select@language{\languagename}%
	\expandafter\ifx\csname date\languagename\endcsname\relax\else
	\if@filesw
	\protected@write\@auxout{}{\string\select@language{\languagename}}%
	\bbl@for\bbl@tempa\BabelContentsFiles{%
		\addtocontents{\bbl@tempa}{\xstring\select@language{\languagename}}}%
	\bbl@usehooks{write}{}%
	\fi
	\fi}
\newcommand{\DeclareLanguageAlias}[2]{%
	\global\@namedef{babel@language@alias@#1}{#2}%
}
\makeatother

\newcommand\varpm{\mathbin{\vcenter{\hbox{%
  \oalign{\hfil$\scriptstyle+$\hfil\cr
          \noalign{\kern-.3ex}
          $\scriptscriptstyle({-})$\cr}%
}}}}
\newcommand\varmp{\mathbin{\vcenter{\hbox{%
  \oalign{$\scriptstyle({+})$\cr
          \noalign{\kern-.3ex}
          \hfil$\scriptscriptstyle-$\hfil\cr}%
}}}}

\DeclareLanguageAlias{en}{english}

\begin{document}

\title{Fast quantum gates for exchange-only qubits using simultaneous exchange pulses}

\author{Irina Heinz}
\affiliation{Department of Physics, University of Konstanz, D-78457 Konstanz, Germany}
\author{Felix Borjans}
\affiliation{Intel Technology Research, Intel Corporation, Hillsboro, OR 97124, USA}
\author{Matthew J. Curry}
\affiliation{Intel Technology Research, Intel Corporation, Hillsboro, OR 97124, USA}
\author{Roza Kotlyar}
\affiliation{Intel Technology Research, Intel Corporation, Hillsboro, OR 97124, USA}
\author{Florian Luthi}
\affiliation{Intel Technology Research, Intel Corporation, Hillsboro, OR 97124, USA}
\author{Mateusz T. Mądzik}
\affiliation{Intel Technology Research, Intel Corporation, Hillsboro, OR 97124, USA}
\author{Fahd A. Mohiyaddin}
\affiliation{Intel Technology Research, Intel Corporation, Hillsboro, OR 97124, USA}
\author{Nathaniel Bishop}
\affiliation{Intel Technology Research, Intel Corporation, Hillsboro, OR 97124, USA}
\author{Guido Burkard}
\affiliation{Department of Physics, University of Konstanz, D-78457 Konstanz, Germany}


\begin{abstract}
    The benefit of exchange-only qubits compared to other spin qubit types is the universal control using only voltage controlled exchange interactions between neighboring spins. As a compromise, qubit operations have to be constructed from non-orthogonal rotation axes of the Bloch sphere and result in rather long pulsing sequences. 
    This paper aims to develop a faster implementation of single-qubit gates using simultaneous exchange pulses and manifests their potential for the construction of two-qubit gates. We introduce pulse sequences in which single-qubit gates could be executed faster and show that subsequences on three spins in two-qubit gates could be implemented in fewer steps. Our findings can particularly speed up gate sequences for realistic idle times between sequential pulses and we show that this advantage increases with more interconnectivity of the quantum dots. We further demonstrate how a phase operation can introduce a relative phase between the computational and some of the leakage states, which can be advantageous for the construction of two-qubit gates. In addition to our theoretical analysis, we experimentally demonstrate and characterize a simultaneous exchange implementation of $X$ rotations in a SiGe quantum dot device  and compare to the state of the art with sequential exchange pulses.
\end{abstract}


\maketitle

\section{Introduction}
In recent years there has been immense progress in spin qubit devices in Si/SiGe heterostructures reaching up to six qubits in a quantum processor \cite{Philips_2022} using Loss-DiVincenzo (LD) qubits \cite{PhysRevA.57.120}. LD qubits are formed by trapping a single electron or hole in a gate-defined quantum dot to use the spin degree as computational unit, the qubit. However, when scaling up LD qubits the delivery of high-power oscillating signals required for qubit rotations causes heating and wiring issues leading to fidelity limitations \cite{PhysRevApplied.19.044078, PhysRevLett.132.067001, kelly2023capacitivecrosstalkgatebaseddispersive, PhysRevX.13.041015}. Thus it is reasonable to investigate qubit types using different operation schemes. In contrast to LD qubits exchange-only (EO) qubits \cite{DiVincenzo_2000} only require one type of interaction, namely the exchange between neighboring spins to realize universal qubit control. Hence, EO qubits utilize baseband pulses and do not need high-power ac fields. 
Moreover, for EO qubits a magnetic field gradient is not required, which represents an advantage regarding scalability. In fact, inhomogeneities of the magnetic field can cause dephasing and leakage in EO qubits \cite{PRXQuantum.2.010347}. 

\begin{figure}[t]
	\centering	\includegraphics[width=0.48\textwidth]{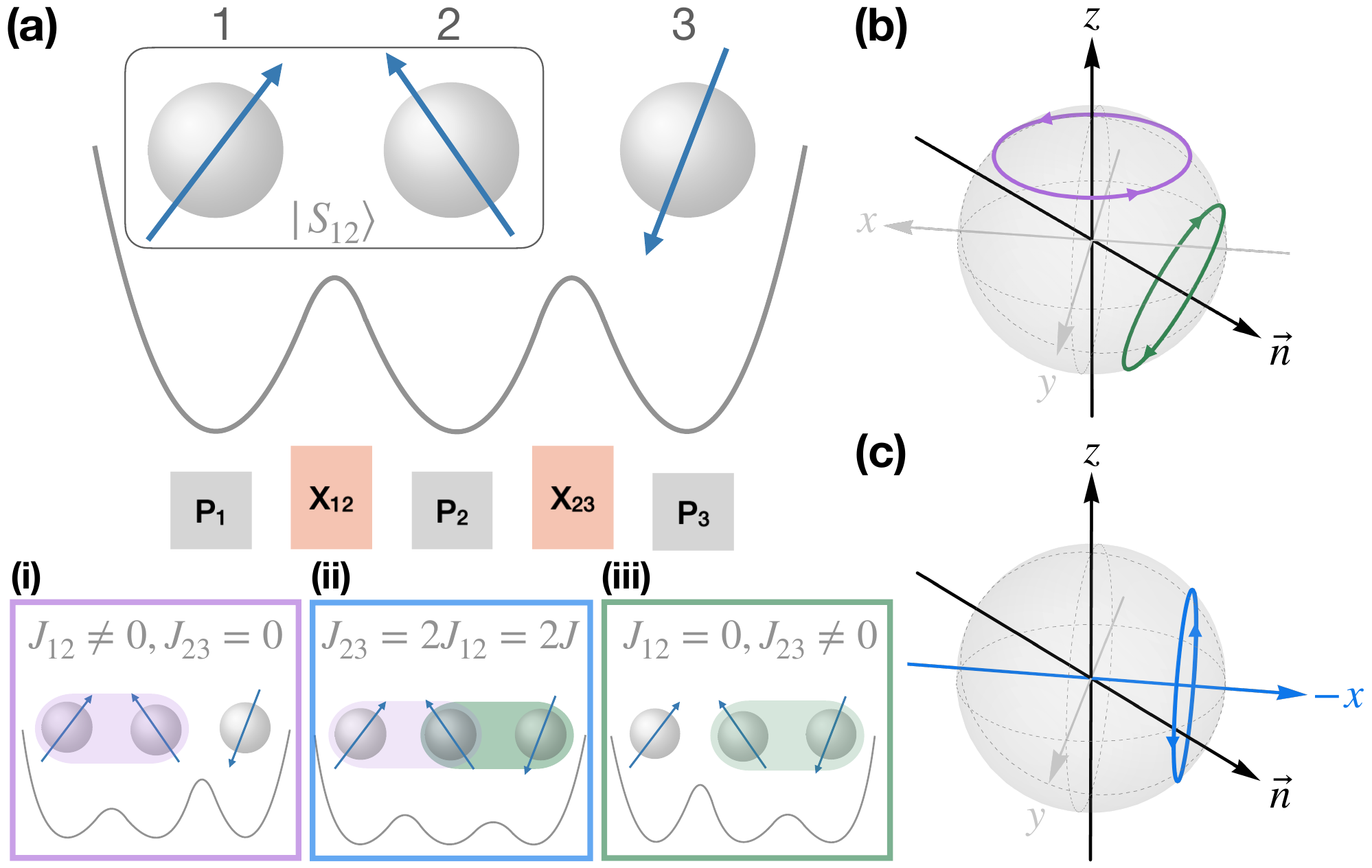}
	\caption{Schematic of an exchange-only qubit encoded in three linearly connected quantum dots in the (1,1,1) charging regime formed by potentials of plunger gates (${\rm P}_i$ of dot $i$) and barrier gates (${\rm X}_{ij}$ between dots $i$ and $j$). (a) Qubits are initialized in the singlet state of spins 1 and 2. Single qubit rotations are realized by switching on the exchange interaction (i) $J_{12}$ for $z$ rotations, (iii) $J_{23}$ for $\vec{n}$ rotations, and (ii) by adjusting the ratio between them the rotation axis can be tuned within the circular sector enclosed by the $z$ axis and $\vec{n}$ to achieve an $x$~rotation. (b) and (c) show rotations on the Bloch sphere color coded respective to the boxes (i)-(iii).}
	\label{Fig:EOqubit}
\end{figure}

In the EO architecture, each qubit is represented by the spins of three electrons hosted in three adjacent quantum dots (Fig.~\ref{Fig:EOqubit}(a)), and quantum gates are generated by pulsing exchange interactions between electrons. Out of the resulting eight-dimensional Hilbert space, four states are computational states and four are leakage states.
Single-qubit gates involve exchange interactions between the electrons belonging to the same qubit and are usually composed of up to four rotations around two different axes \cite{DiVincenzo_2000,NatN8_2019_1Q} (Fig.~\ref{Fig:EOqubit}(b)). Two-qubit gates require exchange interactions between electrons from different qubits. Non-trivial pulse sequences of several exchange pulses populating and depopulating leakage states result in gates such as the CNOT, CZ or SWAP gate \cite{DiVincenzo_2000,FongWandzura, Zeuch1,Zeuch2,Ivanova2024}. With few exceptions, the sequences that are found to result in two-qubit gates, however, assume only commuting exchange pulses to be pulsed at a time. 
Here we investigate the simultaneous application of non-commuting exchange-pulses which will also become relevant for other exchange-based qubit implementations \cite{PhysRevB.93.121410, PhysRevLett.111.050503, bosco2024exchangeonlyspinorbitqubitssilicon}. We obtain quantum gates with fewer pulses, and hence faster executions of gate sequences. We derive conditions on the exchange signals and compare the results to sequential gate operations.

This paper is organized as follows. We introduce the exchange-only qubit in Sec.~\ref{Sec:linear} and theoretically show how single-qubit gates can be implemented using fewer pulses and we experimentally demonstrate the concept of simultaneous exchange pulses in Sec.~\ref{Sec:1Qgates}. We show how a simultaneous gate can perform a direct $X$ rotation by tuning the gate voltages into the appropriate regime and discuss the relative effect of charge noise in comparison to the sequential counterpart. In Sec.~\ref{Sec:2Qgate} we investigate two-qubit gate subsequences and present an operation introducing phases between qubit states and selected leakage states. We extend our description from a linear to an all-to-all connectivity in Sec.~\ref{Sec:all-to-all} and summarize the results for frequently used quantum gates in Tables~\ref{Tab:GateDecomposition} and \ref{Fig:2Qsubsequences}. 

\section{Results} \label{Sec:linear}
To describe exchange-only (EO) qubits in this work we restrict a quantum dot array to be in the (1,1,...) charging regime with one electron in each dot, in which each electron is well described by its spin, and pairwise interactions follow the Heisenberg Hamiltonian
\begin{align}
    H = \sum_{\langle i , j \rangle} J_{ij} (t) \left( \mathbf{S}_i \cdot \mathbf{S}_i -\frac{1}{4} \right) , \label{Eq:Hamiltonian}
\end{align}
where $J_{ij}$ describes the exchange interaction between spin $i$ and $j$. In Appendix~\ref{App:SweetSpot} we derive the exchange interaction $J_{ij}$ from the Hubbard model and show the optimal operation point to be less sensitive to charge noise. An optional global magnetic field can increase the coherence time making transverse fluctuations of the local magnetic field negligible. Global field fluctuations do not have any impact on an EO qubit, since the corresponding operator commutes with the Hamiltonian. In the following we will label the spins (as in Fig.~\ref{Fig:EOqubit}) by numbers from left to right, where only nearest neighbors are connected via exchange interaction.

In the absence of an external magnetic field, and when $J_{ij} = 0$, the EO qubit is in its idle state and all eight energy states are degenerate. Pulsing the exchange interaction causes two of the energy levels to lower and thus pick up a phase relative to the other states. In fact, any operations on one or more qubits are entirely performed by only pulsing the exchange interaction between spins, which will result in a rotation of a subspace of the full Hilbert space.

\subsection{Single-qubit gates}\label{Sec:1Qgates}
\begin{figure*}[ht]
	\centering
	\includegraphics[width=0.65\textwidth]{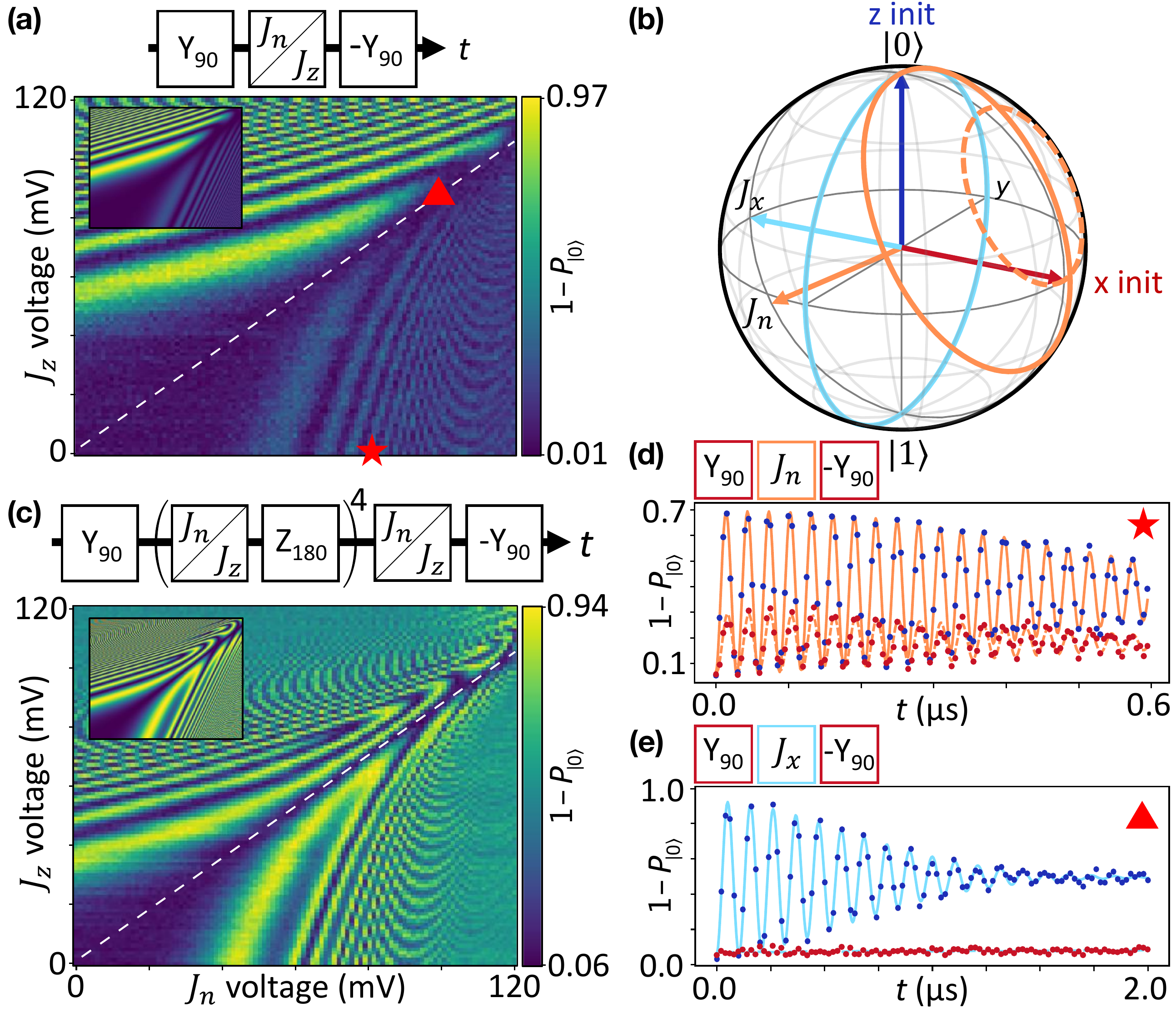}
	\caption{ (a) Experimental fingerpinch plot with $X$ eigenstate prepared. Barriers controlling $J_n$ ($J_{23}$) and $J_z$ ($J_{12}$) exchange are pulsed for a fixed time with amplitudes given by the plot axes. One minus the ground state measurement probability ($1-P_{|0\rangle}$) is plotted for different colors shown on the scale bar to the right. Simulated rotations are plotted in the upper left inlay. (b) Bloch sphere showing $J_n$ (orange) and $J_x$ (blue) rotation axes and rotation trajectories. For $J_n$ rotations, the solid and dashed orange circles represent the trajectory of the singlet state and $X$ eigenstate rotations, respectively. (c) Fingerpinch plot with echoing and repeated exchange pulsing during the pulse sequence. $J_n$ rotations (lower right region) are more visible due to the repeated pulsing and the $Z_{\pi}$ pulses rotating the eigenstate to parts of the Bloch sphere more orthogonal to $J_n$. (d) $J_n$ time-domain rotations for a fixed barrier pulse (red star in (a)) for singlet state (blue points) and $X$ eigenstate (red points). The quality factor of the rotations is 19.9. (e) $J_x$ time-domain rotations with singlet state (blue points) and $X$ eigenstate (red points) at a fixed barrier pulse (red triangle in (a)). The quality factor $Q=8.2$ is lower in this case. The $X$ eigenstate does not rotate, which is expected (red points).}
	\label{Fig:experiment}
\end{figure*}
Instead of the standard spin-up and spin-down basis, the Clebsch-Gordan coefficients are used to transform the single-qubit Hamiltonian to a basis using total spin $S$, its projection $S_z$, and the combined spin on sites 1 and 2, $S_{12}$. We then express the Hamiltonian in terms of the basis states $|S S_z S_{12}\rangle$,
\begin{equation}
    \begin{aligned}
    |0_-\rangle &= \left| \frac{1}{2}, -\frac{1}{2}, 0 \right>, & |1_-\rangle &= \left| \frac{1}{2}, -\frac{1}{2}, 1 \right>,\\
    |0_+\rangle &= \left| \frac{1}{2}, \frac{1}{2}, 0 \right>, & |1_+\rangle &= \left| \frac{1}{2}, \frac{1}{2}, 1 \right>,\\
    |L_1\rangle &= \left| \frac{3}{2}, -\frac{3}{2}, 1 \right>, & |L_2\rangle &= \left| \frac{3}{2}, -\frac{1}{2}, 1 \right>,\\
    |L_3\rangle &= \left| \frac{3}{2}, \frac{1}{2}, 1 \right>, & |L_4\rangle &= \left| \frac{3}{2}, \frac{3}{2}, 1 \right>.
    \label{Eq:basisstates}
\end{aligned}   
\end{equation}
Here $|0_-\rangle$, $|1_-\rangle$, $|0_+\rangle$, $|1_+\rangle$ are the computational basis states with total spin 1/2 and spin projection down and up, respectively, and $|L_1\rangle$, $|L_2\rangle$, $|L_3\rangle$, $|L_4\rangle$ correspond to the four leakage states with total spin 3/2. 

We consider the qubit space for the decoherence-free subsystem with total spin 1/2, i.e., we do not favor one of the two possible spin polarizations. Since exchange coupling preserves the total spin, the Hamiltonian can effectively be written in the $\{|0_{\pm}\rangle, |1_{\pm}\rangle\}$ basis as
\begin{align}
    H(t) = - \frac{1}{2} \left(J_{12} (t)  \sigma_z - \frac{J_{23} (t)}{2} \left(\sigma_z + \sqrt{3}  \sigma_x \right) \right),\label{Eq:1Q}
\end{align}
omitting the inconsequential term 
$\tfrac{1}{2}(J_{12} + J_{23})\openone$ proportional to the identity.
Both blocks $\pm$ are entirely decoupled from the leakage states in the absence of local magnetic noise.

Different from single-spin qubits, EO single-qubit gates need to be constructed from rotations around two non-orthogonal axes, the $z$ axis and the vector $\vec{n}$ (Fig.~\ref{Fig:EOqubit}(b)), using 1 up to 4 pulses \cite{DiVincenzo_2000,NatN8_2019_1Q}. 
A rotation around the $z$ axis is accomplished by only switching on $J_{12}$ ($J_{23}=0$). The time evolution is then given by $R_{z} (\theta_z)=\exp(i \theta_z \sigma_z/2)$ where $\theta_z=\int_0^\tau J_{12}(t) dt$ is the rotation angle. Similarly if turning on only exchange $J_{23}(t)$ ($J_{12}=0$), the time evolution $R_{\vec{n}}(\theta_{\vec{n}}) = \exp(-i \theta_{\vec{n}} (\sqrt{3} \sigma_x + \sigma_z)/ 4)$ with $\theta_{\vec{n}}=\int_0^\tau J_{23}(t) dt$ results in a rotation around the vector $\vec{n}=-(\sqrt{3},0,1)$ (green circle in Fig.~\ref{Fig:EOqubit}(b)). The $X$ and $Y$ gates can be composed by the sequences $X=R_{\vec n} (\theta_1) R_{z} (\theta_2) R_{\vec n} (\theta_1)$ and $Y= R_{z} (\pi)R_{\vec n} (\theta_1) R_{z} (\theta_2) R_{\vec n} (\theta_1)$ with $\theta_1 = \pi -\arctan(\sqrt{8})$ and $\theta_2=\arctan(\sqrt{8})$, respectively \cite{NatN8_2019_1Q}. 

In general, any rotation axis in the $xz$-plane enclosed by the $z$ axis and $\vec{n}$ is possible. Thus a direct rotation around the $x$ axis in the opposite rotational direction (Fig.~\ref{Fig:EOqubit}(c)) can be realized using a simultaneous (or parallel \cite{DiVincenzo_2000}) pulsing of both exchange values with $J_{23} (t)  = 2 J_{12} (t) = 2J(t)$, as the $\sigma_z$ terms cancel in the Hamiltonian in Eq.~\eqref{Eq:1Q}. The time evolution is then given by 
\begin{align}
    R_x (\theta_x) = \exp(-i \theta_x \sigma_x/ 2),
\end{align}
with rotation angle $\theta_x = 2\pi-\int_0^\tau \sqrt{3} J(t) dt$. Note that in this case the Hamiltonian commutes for all times (i.e. $[H(t),H(t')]=0$ for arbitrary $t, t'$). 
When assuming rectangular pulses with maximal experimentally achievable exchange $J_{\rm max}$ the gate time of the sequential $X$ gate is then $\tau_{\rm seq} \approx 5.05/J_{\rm max} + 3\tau_{\rm idle}$, while for the simultaneous $X$ gate the gate time is $\tau_{\rm sim} \approx 3.63/J_{\rm max}+\tau_{\rm idle}$, where $\tau_{\rm idle}$ is the idle time after each pulse \cite{2Qpaper2022}. Consequently, not only is the simultaneous $X$ gate faster compared to its sequential counterpart, by a factor between 1.39 and 3 for realistic pulse times detailed in Table~\ref{Tab:GateDecomposition}, but it also requires two fewer idling times. During these idling times the qubit decoheres due to e.g. magnetic noise and residual exchange. 

These findings indicate that more frequently used single-qubit gates may benefit from the simultaneous pulsing of exchange couplings. Indeed, we find that the $YH$, $X_{\pi/2} = HSH$ and $X_{-\pi/2} = HS^{\dagger}H$ gates, where $S={\rm diag}(1,i)$ and $H$ is the Hadamard gate, can be implemented using one simultaneous pulse step instead of three sequential ones. Furthermore, most gates in Table~\ref{Tab:GateDecomposition} can be implemented using two pulses instead of three, which results in a decrease of the average gate time by~$\approx 18\%$.


To verify and characterize the operation of simultaneous exchange, a SiGe quantum dot device fabricated by Intel \cite{Neyens_2024} is tuned up as an exchange-only qubit \cite{doi:10.1126/sciadv.1500214}. The electron spins are hosted in a Si/Si$_{0.7}$Ge$_{0.3}$ heterostructure grown on a 300-mm Si wafer. For the experiment the 12-dot Intel Tunnel Falls device that can host up to 12 electrons in a linear array was used to operate a single exchange only qubit \cite{doi:10.1021/acs.nanolett.4c05205}. The qubit is encoded across three quantum dots underneath metal “plunger” gates with “barrier” gates in between dots to control the exchange. The qubit is controlled via baseband pulses on the barrier gates and readout is performed using Pauli spin blockade with signal amplified by SiGe heterojunction bipolar transistor (HBT) cryogenic amplifiers \cite{doi:10.1126/science.1116955, Curry2019}.

To reduce sensitivity to charge noise and calibrate exchange, “fingerprint” plots \cite{PhysRevLett.116.110402} are acquired for the “$J_z$” and “$J_n$” control axes, corresponding to $J_{12}$ and $J_{23}$, respectively, where $J_n$ points down 120 degrees from the $z$-axis, analogous to Fig.~\ref{Fig:EOqubit}(b). For that the exchange interaction is varied as a function of barrier voltage and detuning, leading to the typical appearance of a fingerprint. In fact, for a fixed time of 100 ns, the barrier gate bias is pulsed to larger values while the voltage detuning between two surrounding plunger gates is pulsed from negative to positive values. $J_n$ fingerprint plots can be taken after preparing a singlet state, but $J_z$ fingerprints must have a pre- and post-$J_n$ rotation applied which brings the state to and from the $xy$-plane of the Bloch sphere, respectively.

A typical fingerprint plot uses two plungers and one barrier to increase only one exchange term and calibrate one rotation axis, which can then be used for serial exchange control. To increase both exchange terms simultaneously, the two barrier gates are pulsed together for a fixed time, and $1-P_{|0\rangle}$ is plotted in the “fingerpinch” plot in Figure~\ref{Fig:experiment}(a), where $P_{|0\rangle}$ is the probability of measuring the ground state $|0_\pm\rangle$. The Pauli spin blockade readout only detects whether the two neighboring spins 1 and 2 in Fig.~\ref{Fig:EOqubit}(a) are in a singlet state or not, entirely neglecting the spin projection $S_z$. If a singlet state is prepared, it will only be rotated by the $J_n$ exchange, therefore, we prepare an $X$ eigenstate using a $Y_{\pi/2}$ prerotation, which uses alternating $J_n$ and $J_z$ exchange pulses \cite{NatN8_2019_1Q}. The $Y_{\pi/2}$ pulse is composed of one $J_n$ pulse that rotates the singlet state to the equator of the Bloch sphere followed by a $J_z$ pulse which rotates the state to the $X$ eigenaxis. The $X$ eigenstate can be rotated by either $J_n$ (lower right region of plot) or $J_z$ (upper left region). Additionally, the $X$ eigenstate will not be rotated when the total exchange points along the $x$-axis, therefore the darker, linear region of the upper right defines the “$J_x$” regime and is indicated with a white dashed line. Theoretical calculations of the expected rotation outcomes are plotted in the upper left and agree well with the data. In the simulated plots, a symmetric -37\% cross capacitance was added which was obtained by a qualitative fit to the data.

A modified fingerpinch plot is acquired and plotted in Figure~\ref{Fig:experiment}(c), where alternating $Z_{\pi}$ gates and exchange pulses are applied to echo the state and extend the rotation visibility, particularly in the $J_n$ regime (lower right region of plot). Here, the $J_x$ axis highlighted by the white dashed line is more visible throughout the plot with theoretical predictions again in agreement. Figure~\ref{Fig:experiment}(d) shows time-domain data for $J_n$ rotations, where the barrier is pulsed to a fixed value shown by the red star in figure~\ref{Fig:experiment}(a). Either a singlet state (blue data points) or an $X$ eigenstate (red points) is prepared as depicted as a blue and red initial state on the Bloch sphere in Fig.~\ref{Fig:experiment}(b). The frequency of the rotations is $f=33.9$~MHz with a quality factor of $Q=19.9$. The quality factor, $Q$, is defined as the characteristic decay time of an exponentially decaying sinusoid fit to the rotation data multiplied by the frequency of the rotations. Similarly, Figure~\ref{Fig:experiment}(e) shows the $J_x$ time-domain data, taken with barriers pulsed to the upper right region of the fingerpinch plot (red triangle point in Figure~\ref{Fig:experiment}(a)). The frequency is $f=9.5$~MHz and the quality factor is a lower value $Q=8.2$. As expected, $J_x$ does not rotate an $X$ eigenstate (red points). The rotations corresponding to $J_n$ and $J_x$ are shown in Fig.~\ref{Fig:experiment}(b) in orange and light blue, respectively.

The lower quality factor of the $J_x$ rotations may be due to multiple effects. Since the system now forms two charge dipoles instead of one, any charge noise present in the immediate area may couple in to a greater degree. Additional optimizations to tune into a “sweet spot” to minimize charge noise may be possible. Follow-up experiments include pulsing the detuning axis of the two outer plungers as the two barrier gates are pulsed to greater values with different offsets.

\begin{figure}[ht]
	\centering\includegraphics[width=0.46\textwidth]{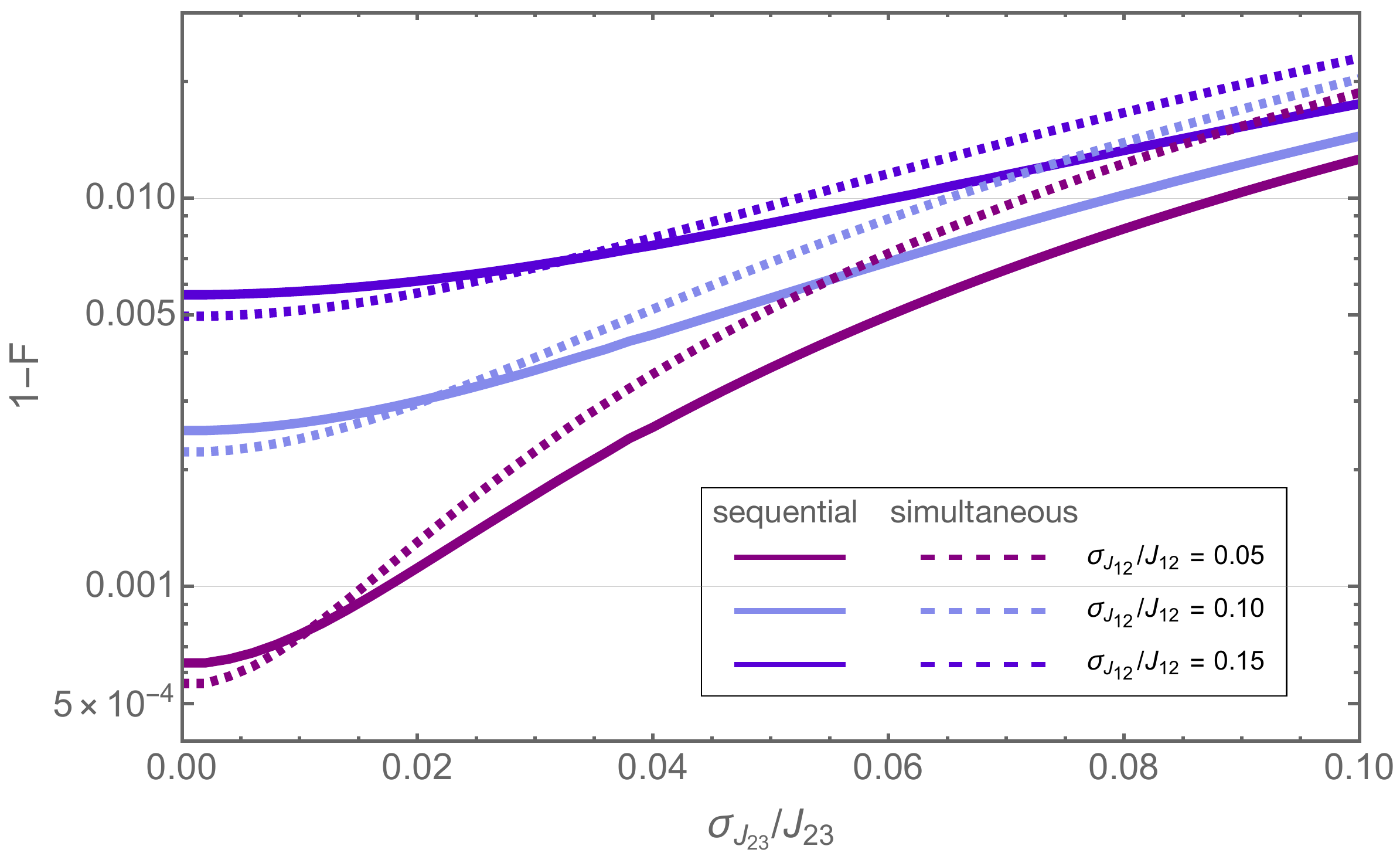}
	\caption{Calculated $X$ gate fidelity suffering from quasi-static exchange fluctuations with standard deviations $\sigma_{J_{12}}$ and $\sigma_{J_{23}}$. We compare the sequential and simultaneous $X$ gate (with linear connectivity) for various combinations. Both, the sequential and the simultaneous pulse gate fidelity strongly depend on the fluctuations in $J_{23}$, whereas the fluctuations in $J_{12}$ are rather negligible.}
	\label{Fig:noise}
\end{figure}

Charge noise enters in the Hamiltonian \eqref{Eq:Hamiltonian} as fluctuations in the exchange interaction. A simple estimation for the fidelity is to assume quasi-static exchange fluctuations on each exchange value separately, i.e. $J_{ij}\rightarrow J_{ij}+\delta J_{ij}$, where $\delta J_{ij}$ is a Gaussian-distributed exchange fluctuation with standard deviation $\sigma_{J_{ij}}$. In Fig.~\ref{Fig:noise} we compare the sequential (solid lines) and simultaneous (dashed lines) $X$ gate with linear connectivity for various $\sigma_{J_{12}}/J_{12}$, where $\sigma_{J_{12}}$ is the standard deviation for fluctuations in $J_{12}$. We find that both, the sequential and the simultaneous pulse gate fidelities have a similar behaviour with a stronger dependence on the fluctuations in $J_{23}$ than in $J_{12}$. For the sequential gate this can be explained by the larger overall rotation angle around $\vec{n}$ induced by $J_{23}$ in the $X$ gate construction, whereas for the simultaneous gate it is due to the fact that $J_{23} = 2 J_{12}$. Any deviation from this ratio results in a tilt of the rotation axis and can result in an additional over- or underrotation. For larger $\sigma_{J_{23}}/J_{23}$ we find a slightly better performance for the sequential gate, however, for biased noise in the different exchange interactions with, e.g., $\sigma_{J_{12}}/J_{12} = 10$\% and $\sigma_{J_{23}}/J_{23} \leq 2$\%, we even find a fidelity improvement when using the simultaneous pulses.
Typical exchange fluctuations $\sigma_{J_{ij}}/J_{ij}$ in the experiment in Fig.~\ref{Fig:experiment} correspond to 0.01 to 0.03. These values were estimated from simulations of the time dynamics in a double quantum dot which assume that $J$ has a standard deviation of $\sigma_J$ leading to the relation $\sigma_J/J \approx 0.22/Q$ (so for $Q = 8.2 - 19.9$ we find $\sigma_J/J = 0.01 - 0.03$).

\subsection{Two-qubit gate subsequences}\label{Sec:2Qgate}
\begin{figure*}[ht]
	\centering
	\includegraphics[width=0.98\textwidth]{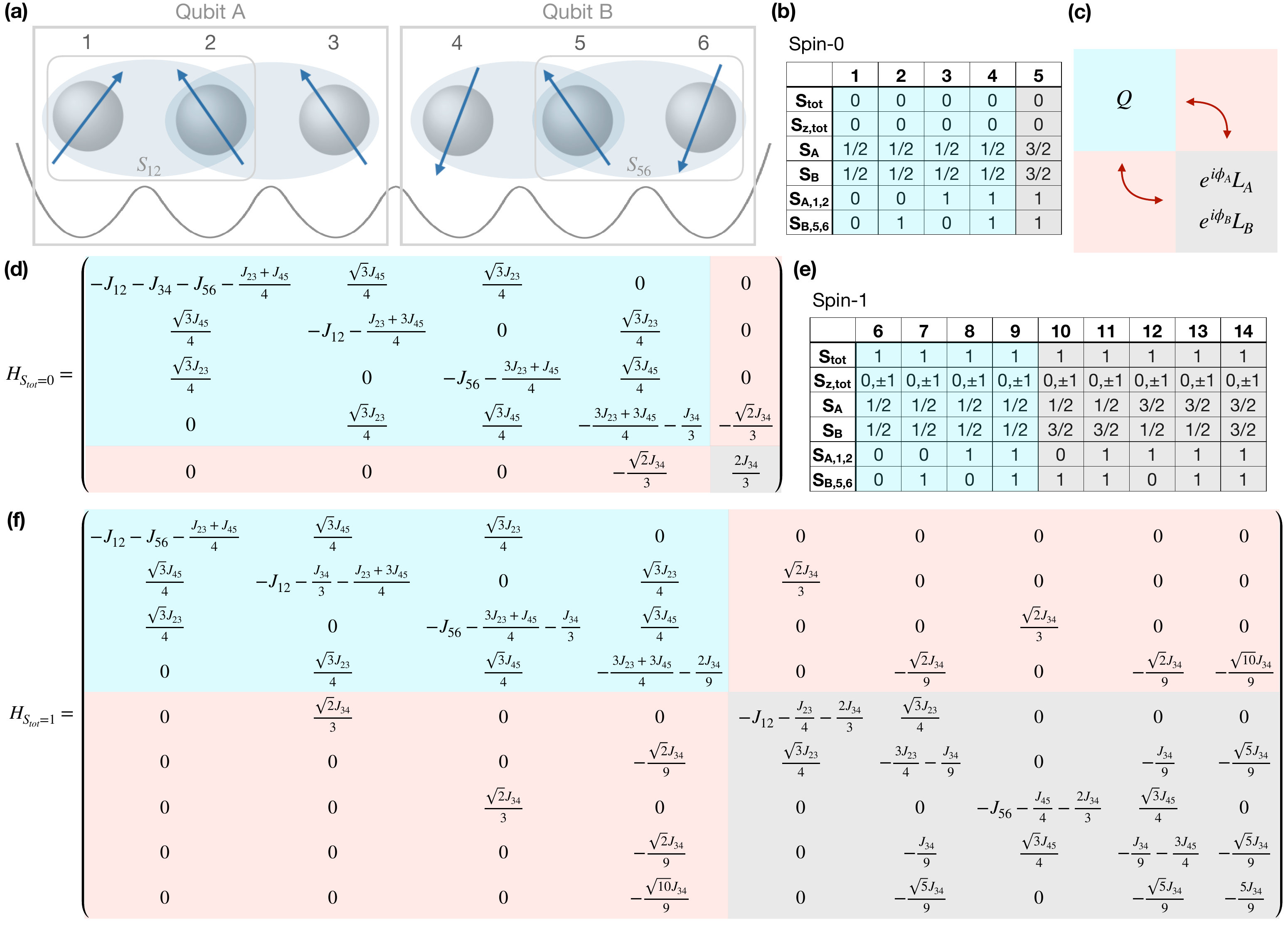}
	\caption{(a) Schematic of a chain of 6 spins, where the outer spin pairs are initialized in singlets. Spins 1-3 form qubit $A$ and spins 4-6 form qubit $B$. (b) and (e) Total spin basis with total spin $S_{\rm tot}$, spin $z$-projection $S_{z,{\rm tot}}$, total spin of qubit $A$ and $B$ $S_{A}$ and $S_B$ and total spin of the outer two spins $S_{A,1,2}$ and $S_{B,5,6}$ for the $S_{\rm tot}=0,1$ subspaces. The computational states of the $S_{\rm tot} = 0$ ($S_{\rm tot} = 1$) block are 1,2,3 and 4 (6,7,8 and 9). The $S_{\rm tot} = 2,3$ subspaces are completely outside of the computational space. (c) Schematic of qubits space in light blue and leakage space in light gray. Non-zero entries in the red blocks would couple qubit states to leakage states (red double arrows). Instead, when using simultaneous exchange pulses as in Eqs.~\eqref{Eq:phase-correction-linearQAS0}-\eqref{Eq:phase-correction-linearQBS1}, relative phases $\phi_A$ and $\phi_B$ between some of the qubit and leakage states are introduced, visualized by $L_A$ and $L_B$, respectively. (d) Two-qubit Hamiltonian of the $S_{\rm tot}=0$ subspace written in the total spin basis vectors from (b). (f) Two-qubit Hamiltonian of the $S_{\rm tot}=1$ subspace in the total spin basis vectors from (c), neglecting $S_{z,\rm tot}$.}
	\label{Fig:2Qbasis}
\end{figure*}

To construct two-qubit gates for two qubits $A$ and $B$, it is advantageous to use the total spin basis of the entire six spin system $|S_{\rm tot} S_{z,\rm tot} S_{A} S_{B} S_{A,1,2} S_{B,1,2}\rangle$. Since $S_{\rm tot}$ and $S_{z,\rm tot}$ are preserved under exchange interactions, blocks with different total spins are separable as they do not couple to each other \cite{DiVincenzo_2000, FongWandzura}. As shown in Fig.~\ref{Fig:2Qbasis}(a) we label the chain of six spins from left to right with numbers and assume the outer two spins of each triple to be initialized in a singlet configuration as described in Ref.~\cite{2Qpaper2022}. We label the left three spins as qubit~$A$ and the right three spins as qubit~$B$. For this configuration we list the possible basis states in the total spin basis of the $S_{\rm tot} = 0$ and 1 subspace in Fig.~\ref{Fig:2Qbasis}(b) and (e), respectively. Each qubit, $A$ and $B$, is operated in the single-qubit spin $S_A = S_B=1/2$ subspace, thus only the total spin subspaces $S_{\rm tot}=0$ and $S_{\rm tot}=1$ cover the computational space. For each block the Hamiltonian is schematically shown in Fig.~\ref{Fig:2Qbasis}(c). The computational subspace is colored in light blue and the leakage states in gray. The Hamiltonian in Eq.~\eqref{Eq:Hamiltonian} can be written in the $S_{\rm tot} = 0,1$ blocks as shown in Fig.~\ref{Fig:2Qbasis}(d) and \ref{Fig:2Qbasis}(f), respectively.

\begin{figure}[b]
	\centering\includegraphics[width=0.48\textwidth]{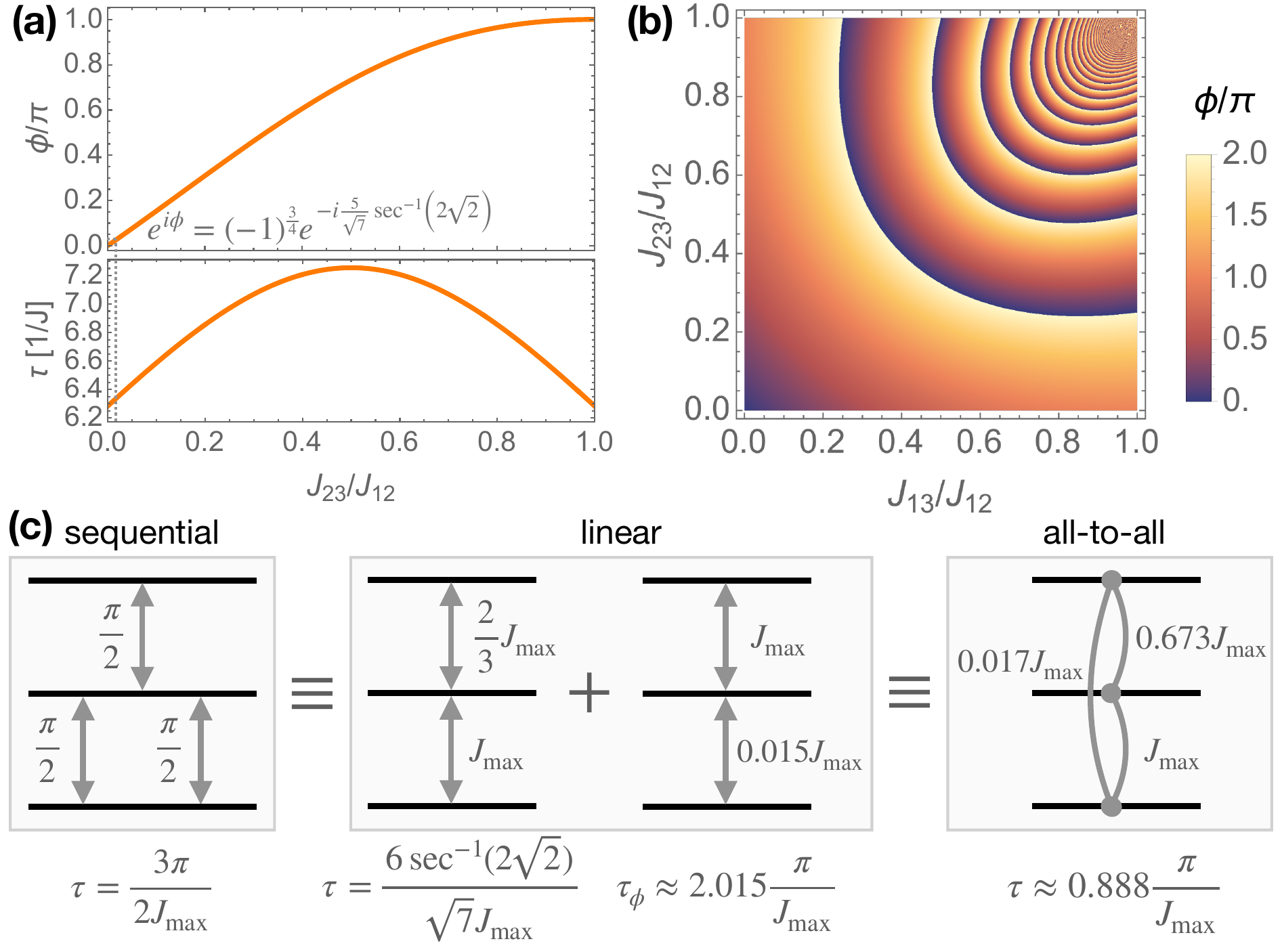}
	\caption{(a) Correction phase $\phi$ as described in Eqs.~\eqref{Eq:phase-correction-linearQAS0}--\eqref{Eq:phase-correction-linearQAS1} for qubit $A$ depending on the ratio $J_{23}/J_{12}$ where $J_{13}=0$ (linear connectivity) and corresponding gate time $\tau$ in units of $1/J_{\rm max}$. The vertical gray dashed line shows the required phase in example (c) with linearly connected spins. (b) Correction phase with all-to-all connectivity can achieve arbitrary angles $\phi$ depending on the respective exchange ratios. (c) Equivalent pulses for sequential and simultaneous pulsing with linear and all-to-all connectivity.}
	\label{Fig:Phases}
\end{figure}

In this basis one can construct sub-sequences containing rotations in subspaces of the full Hilbert space as in Refs. \cite{Zeuch1, Zeuch2, Zeuch_2020} or apply e.g. genetic algorithms to search for a pulse sequence overall resulting in 
\begin{align}
    U_{S_{\rm tot}=0,1} = \begin{pmatrix}U^{\rm 2Q gate} & 0\\ 0 & U^{L}_{S_{\rm tot}=0,1} \end{pmatrix}.
\end{align}
Here $U^{\rm 2Q gate}$ corresponds to the unitary describing the two-qubit gate operation in the subspace spanned by the first four basis states (1-4 and 6-9 respectively) neglecting $S_{z,\rm tot}$, whereas $U^{L}_{S_{\rm tot}}$ can be an arbitrary unitary acting only within the non-computational leakage states (basis states 5 and 10-14, respectively). For subspaces $S_{\rm tot} = 2$ and $S_{\rm tot} = 3$ the unitaries $U_{S_{\rm tot} = 2}$ and $U_{S_{\rm tot} = 3}$ can be arbitrary in general. Although the leakage states lie outside the computational space, it is necessary to take them into account when constructing gate sequences; also, the leakage states can be beneficial for reducing leakage errors. Examples for leakage-controlled CZ (LCCZ) and CNOT (LC-CNOT) gates are given in Ref.~\cite{2Qpaper2022}.

Since we are seeking shorter gate sequences, we investigate combinations of subsequences in some of the relevant two-qubit gates such as CZ, CNOT and SWAP. A trivial but useful fact in larger sequences is that if the first and third rotation are by an angle of $\pi$, $\theta_1 = \theta_3 = \pi$, the rotation sequence can be mirrored, i.e. $R_{\vec n} (\pi) R_{z} (\theta_2) R_{\vec n} (\pi) = R_{z} (\pi) R_{\vec n} (\theta_2) R_{z} (\pi)$. As the $\pi$-rotations swap the spins in such a way, that the outer spins interact with each other and then bring them back into the initial order, it does not matter which spins, i.e. 1 and 2 or 2 and 3, are swapped to achieve this. 
In general, non-adjacent exchange interactions can be pulsed at the same time as their Hamiltonians commute.  While two subsequent adjacent exchange pulses cannot be replaced by a single pulse, three exchange pulses on three spins can be compressed in this way if the resulting rotation within the respective subspace is around a feasible axis (i.e., any axis between the z axis and $\hat{n}$). Therefore, we focus on subsequences that are performed on three spins. Note that if these spins belong to the same qubit these ultimately are single-qubit gates within a larger sequence. Furthermore, a necessary but not sufficient condition for replacing a pulse sequence with three pulses (two alternating exchange interaction pulses $R_{\vec n} (\theta_3) R_{z} (\theta_2) R_{\vec n} (\theta_1)$ or $R_{z} (\theta_3) R_{\vec n} (\theta_2) R_{z} (\theta_1)$)  is that the first and third rotation angles are the same, i.e. $\theta_3=\theta_1$.

In the following we discuss the construction of simultaneous gates to replace three-pulse sequences in two qubit gates. For that we (i) identify suitable subsequences, (ii) describe the construction only in a single-qubit computational subspace to find the respective gate, and (iii) correct for any relative phase to the leakage space. 

From several sequences in Ref.~\cite{2Qpaper2022} we first extract those three-pulse combinations, for which first and third rotation angles are equal and for which we can find alternative simultaneous pulses if they were applied only on a single qubit. The original sequence is shown in the first column of Table~\ref{Fig:2Qsubsequences}. For all of these sequences we can find a simultaneous exchange pulse, fulfilling the given conditions, such that the computational space unitary equals the 3-pulse sequence (shown in the left half of the second column). For actual single-qubit gates this is sufficient. However, since during two-qubit gates the purposeful population and depopulation of leakage states is crucial, the relative phases between the total basis states 1-14 in Fig.~\ref{Fig:2Qbasis}(b) and Fig.~\ref{Fig:2Qbasis}(e) are of importance. We observe that during simultaneous pulsing of the exchange values, the relative phase $\varphi$ (given in the third column in Tab.~\ref{Fig:2Qsubsequences}) between some of the states changes compared to the 3-pulse sequence. For instance, if the simultaneous pulse is executed on qubit $A$, in the $S_{\rm tot} =0$ block the states 1-4 in Fig.~\ref{Fig:2Qbasis}(b) acquire a relative phase and in the $S_{\rm tot} =1$ block states 6-11 in Fig.~\ref{Fig:2Qbasis}(e) acquire the same relative phase, while states 5 and 12-14 do not. Thus, to obtain the exact same gate sequence one has to account for this phase in a second pulse. We add a subsequent simultaneous exchange pulse with $J_{12},J_{23} \neq 0$ and diagonalize the time evolution by choosing a pulse duration $\tau_A= \frac{2 \pi}{\sqrt{J_{12}^2-J_{12}J_{23}+J_{23}^2}}$ to obtain unitaries that  cancel out relative phases between the respective  computational and leakage states
\begin{align}
    U_{S_{\rm tot} =0}^{\phi_A} &= -e^{i\phi_A} {\rm diag} (1, 1, 1, 1, -e^{i\phi_A}), \label{Eq:phase-correction-linearQAS0} \\
    U_{S_{\rm tot} =1}^{\phi_A} &= -e^{i\phi_A} {\rm diag} (1, 1, 1, 1, 1, 1, -e^{i\phi_A}, -e^{i\phi_A}, -e^{i\phi_A}), \label{Eq:phase-correction-linearQAS1}
\end{align}
where $\phi_A=\frac{J_{12}+J_{23}}{2} \tau_A$. Analogously, if qubit $B$ is operated, i.e. $J_{45},J_{56} \neq 0$, we obtain
\begin{align}
    U_{S_{\rm tot} =0}^{\phi_B} &= -e^{i\phi_B} {\rm diag} (1, 1, 1, 1, -e^{i\phi_B}), \label{Eq:phase-correction-linearQBS0}\\
    U_{S_{\rm tot} =1}^{\phi_B} &= -e^{i\phi_B} {\rm diag} (1, 1, 1, 1, -e^{i\phi_B}, -e^{i\phi_B}, 1, 1, -e^{i\phi_B}) ,\label{Eq:phase-correction-linearQBS1}
\end{align}
with $\phi_B=\frac{J_{12}+J_{23}}{2} \tau_B$ and $\tau_B= \frac{2 \pi}{\sqrt{J_{45}^2-J_{45}J_{56}+J_{56}^2}}$. We note that as long as $J_{34}=0$ the Hamiltonians of qubit $A$ and $B$ commute and thus both operations can be pulsed simultaneously, if necessary. From the expressions for $\phi_A$ and $\phi_B$ we can obtain the possible correction angles in a single drive which lie between $0$ and $\pi$ as depicted in the upper plot in Fig.~\ref{Fig:Phases}(a) for a varying ratio of $J_{23}/J_{12}$. The lower plot shows the corresponding pulse time. Consequently, we find a two-step simultaneous exchange pulse sequence to replace a three-step single pulse sequence, in case the correction phase lies in the respective phase range. As an example in Fig.~\ref{Fig:Phases}(c) the $\pi/2-\pi/2-\pi/2$ sequence is shown, which can be replaced by two simultaneous pulses. The total pulse time $\tau^{\rm sim}_{\rm tot}$ for the simultaneous case exceeds the total pulse time $\tau^{\rm seq}_{\rm tot}$ of the sequential counterpart, $2.887 \pi/J_{\rm max} > 1.5 \pi/J_{\rm max}$. Nevertheless, due to finite fall times of the barrier voltages, an idle time $\tau_{\rm idle}$ is usually required. For the sequential gate this yields a total gate time $\tau^{\rm seq}_{\rm tot} + 3\tau_{\rm idle}$ while for the simultaneous pulses the total gate time is $\tau^{\rm sim}_{\rm tot}+2\tau_{\rm idle}$. Ultimately, for a long idle time $\tau_{\rm idle}>1.387\pi/J_{\rm max}$, the simultaneous pulsing becomes beneficial compared to the sequential state of the art. Furthermore, by taking into account the correction phase in a second pulse, the three-pulse sequences in Tab.~\ref{Fig:2Qsubsequences} of any neighboring exchange pair can be replaced. In Appendix~\ref{App:2Qgates} we exemplarily replace the three-step sequences with the two-step simultaneous pulses in the Fong-Wandzura CZ gate sequence, calculate the gate times, and estimate the gate performance.
However, there we do not find an overall time improvement, since the circuit was designed and optimized for the brick structure, where only commuting exchange interactions are allowed to be driven simultaneously. Instead, our result shows that there is room for improvement of the currently shortest sequences, an insight that may trigger the exploration of new two-qubit pulse sequences optimized for simultaneous pulsing.

\subsection{All-to-all connectivity} \label{Sec:all-to-all}
\begin{figure}[ht]
	\centering
	\includegraphics[width=0.38\textwidth]{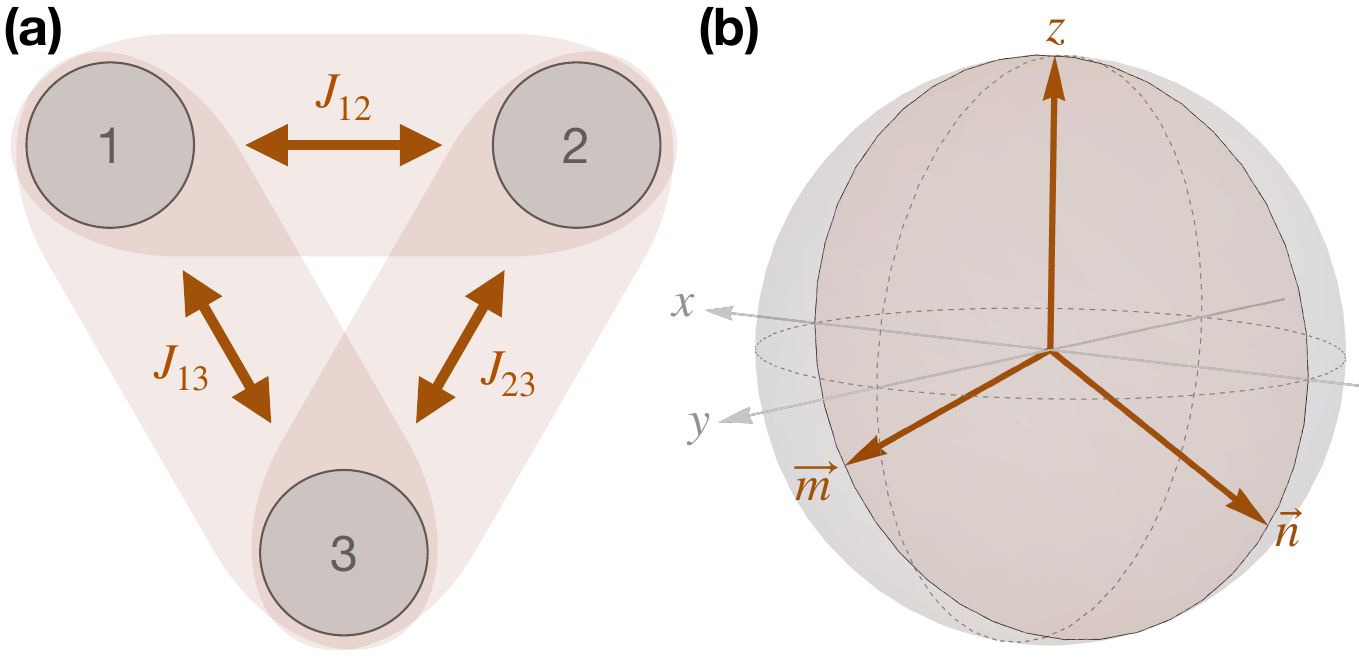}
	\caption{(a) Schematic of an exchange-only qubit with all-to-all connectivity and (b) the corresponding Bloch sphere. Any rotation axis within the plane spanned by the vectors $\vec{n}=-(\sqrt{3},0,1)/2$ and $\vec{m}=(\sqrt{3},0,-1)/2$ is possible by adjusting the ratio between exchange interactions $J_{12}$, $J_{23}$, and $J_{13}$.}
	\label{Fig:All-2-all}
\end{figure}
Although, for the most part, exchange-only qubits were demonstrated in 1D qubit chains, we  point out that in future 2D layouts, the connectivity of exchange-only qubits might vary \cite{Setiawan_2014}. Indeed, recently a triangular exchange-only spin qubit was demonstrated \cite{acuna2024coherent}. Additionally, the simultaneous pulsing of exchange interactions could give rise to a super-exchange and many-body interactions between next-nearest neighbors \cite{Braakman_2013,PhysRevLett.92.077903}. Thus, we consider the Hamiltonian in Eq.~\eqref{Eq:Hamiltonian} with three spins and all-to-all connection ($J_{12}$, $J_{23}$ and $J_{13}$) as depicted in Fig.~\ref{Fig:All-2-all}(a). Again, we can separate the Hamiltonian in the basis of $\{|0_{\pm}\rangle, |1_{\pm}\rangle\}$ into
\begin{align}
    H(t) = \frac{\sqrt{3}}{4} (J_{23}-J_{13}) \sigma_x + \frac{1}{4} (-2J_{12} + J_{23} + J_{13}) \sigma_z,
\end{align}
and a term proportional to the identity, leading to a phase of $e^{i (J_{12} + J_{23} + J_{13})t/2}$ between qubit and leakage space. Similar to the linear connectivity case the insensitivity to voltage fluctuations in the dot potential \cite{PhysRevB.93.121410, PhysRevB.94.165411} can be derived from a Hubbard model as shown in the Appendix~\ref{App:SweetSpot}.

In  case all interactions are independently tunable, the previously discussed pulsing schemes still apply. Furthermore, all rotation axes lying in the $xz$-plane are possible in a single simultaneous pulse \cite{shim2013singlequbitgatesstepsrotation}, as depicted in Fig.~\ref{Fig:All-2-all}(b). The two vectors $\vec{n}=-(\sqrt{3},0,1)$ and $\vec{m}=(\sqrt{3},0,-1)$ correspond to $J_{23}$ and $J_{13}$, respectively. In particular a single-pulse implementation of a rotation around the $x$-axis requires $J_{12}=(J_{23} + J_{13})/2$. We find a rotation around the vector $(1,0,0)$ for $J_{23} > J_{13}$ and around the vector $(-1,0,0)$ for $J_{23} < J_{13}$. Thus $x$ rotations around an angle $\theta \in (0,\pi)$ can be realized fastest with $J_{13}=0$ and around $\theta \in (\pi,2\pi)$ can be realized fastest with $J_{23}=0$. Moreover, a frequently used gate in quantum algorithms, the Hadamard gate, can be realized using one simultaneous pulse with a gate time comparable to the $Z$ and $X$ gate. We summarize and compare simultaneous single-qubit gates to sequential pulses and linearly connected qubit architectures in Tab.~\ref{Tab:GateDecomposition}. We point out that whereas excluding the identity the average gate time for sequential pulses is 73.420~ns, for simultaneous pulses in linearly connected quantum dots it is 60.010~ns, while for all-to-all connectivity it reduces to 44.136~ns. Here we assumed rectangular exchange pulses with maximal exchange value $J_{\rm max}=100$~MHz each. As the gate time reduces with simultaneous pulsing the number of applied quantum gates can be increased until the qubit loses its coherence.

Analogously to the linear case, we again consider the subsequences in two-qubit gates given in Tab.~\ref{Fig:2Qsubsequences} and find single-pulse implementations. Since all-to-all connectivity gives an additional degree of freedom it is possible to compensate for phases within one pulse. In Fig.~\ref{Fig:Phases}(c) the single-pulse gate with all-to-all connectivity equivalent to a $\pi/2-\pi/2-\pi/2$ sequence is showcased. In Appendix~\ref{App:2Qgates} we show how three-step sequences can be replaced by simultaneous pulses in the Fong-Wandzura CZ gate.

Furthermore, not only can these selected subsequences be achieved when using simultaneous pulsing with all-to-all connected spins, but we can also engineer the relative phase $\phi$ between qubit and leakage states by diagonalizing the time evolution using $\tau=2\pi/\sqrt{J_{12}^2 + J_{13}^2  + J_{23}^2 - J_{13} J_{23} - J_{12} J_{13} - J_{12} J_{23}}$. Then we obtain a time evolution as in Eqs.~\eqref{Eq:phase-correction-linearQAS0} and \eqref{Eq:phase-correction-linearQAS1} where the phase $\phi_A= \phi = (J_{12} + J_{13} + J_{23}) \tau/2$ can take any possible value $0\leq \phi<2\pi$. In Fig.~\ref{Fig:Phases}(b) $\phi$ is calculated depending on the ratios $J_{13}/J_{12}$ and $J_{23}/J_{12}$.

\section{Conclusions}
In this paper we have shown an alternative pulsing scheme to operate exchange-only qubits using two neighboring exchange pulses simultaneously. We have found conditions for faster single-qubit gates with fewer pulses and compared the linear and all-to-all connected arrangement, and found that the latter offers more flexibility and thus even faster gates. Increasing the connectivity between spins comes along with challenges such as crosstalk on the control gate electrodes that can be calibrated using virtual gates \cite{PhysRevApplied.13.054018, chadwick2024shorttwoqubitpulsesequences}. Also the number of residual exchange interactions present during idling times increases. On the other hand, the number of control pulses necessary for a specific gate decreases with increased connectivity. This reduces the number of idling times between two pulses.

By investigating reoccurring subroutines in two-qubit sequences, we provide gate implementations in fewer steps and propose to search for a construction of two-qubit gates using subroutines of simultaneous exchange pulses. We further show that with simultaneous pulses we can introduce relative phases between qubit and various leakage states, which can also be used as subroutines when constructing two-qubit gates sequences. We expect this construction to become advantageous for two-qubit gates and possibly even for error mitigation schemes to account for leakage errors.

We also prove the concept with an experimental realization of a direct $X$ gate using simultaneous pulses and characterize the operation regime for the two exchange interactions. Although we believe the experiment is limited by charge noise, we theoretically predict a charge noise regime in which the simultaneous $X$ gate performs better than the state-of-the-art construction utilizing three single exchange pulses. In addition to the presence of noise and decoherence in exchange-only qubits, some remaining issues with our proposed scheme are the precise calibration and voltage control in the device, which needs to be improved in future experiments.
We conclude that our proposed concept for the pulse construction with simultaneous exchange interactions offers a great potential to reduce the number of pulses for faster quantum gates in exchange-only qubits being compatible with other qubit implementations.

\begin{table*}[ht]
	\centering
	\includegraphics[width=0.82\textwidth]{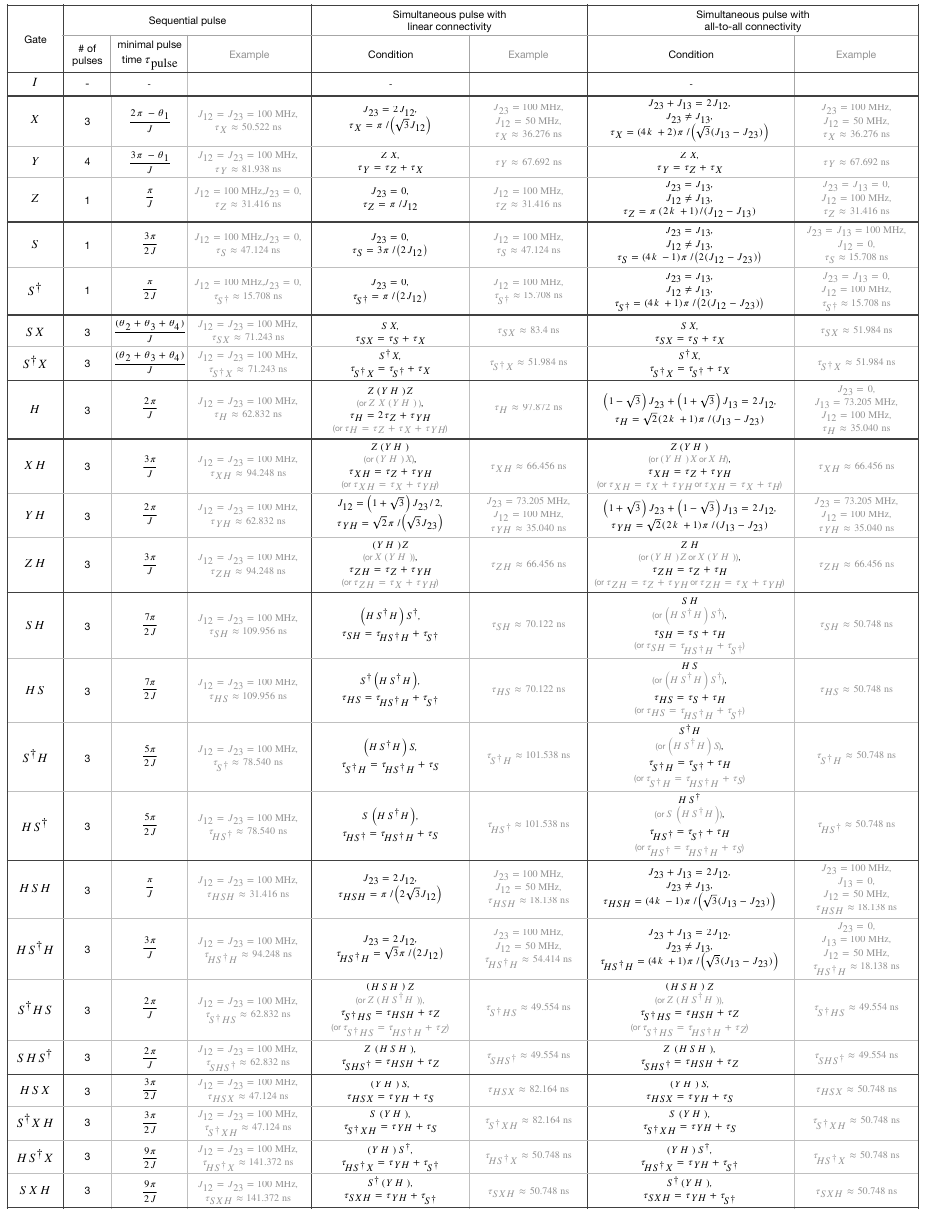}
	\caption{Construction of single qubit gates listed in Ref.~\cite{NatN8_2019_1Q} using simultaneous pulses with linear and all-to-all connectivity. For comparison the number of pulses and gate times using sequential pulses are shown, where $\theta_1 =\arctan(\sqrt{8})$, $\theta_2 =\pi - \arctan(\sqrt{5}/2)$, $\theta_3 \approx 1.305$ and $\theta_4 \approx 3.519$. Explicit examples are shown for gates that can be replaced by a single simultaneous pulse, whereas the others are constructed from two or three pulses. Neglecting the idle times between pulses the average gate time is 73.420~ns for the sequential pulses, 60.010~ns for the simultaneous pulses with linear connectivity and 44.136~ns for the simultaneous pulses with all-to-all connectivity. Here we did not count the identity operation as a gate and assumed a maximal experimentally feasible exchange interaction to be $J_{\rm max} = 100$~MHz.}
	\label{Tab:GateDecomposition}
\end{table*}

\begin{table*}[ht]
	\centering
	\includegraphics[width=1\textwidth]{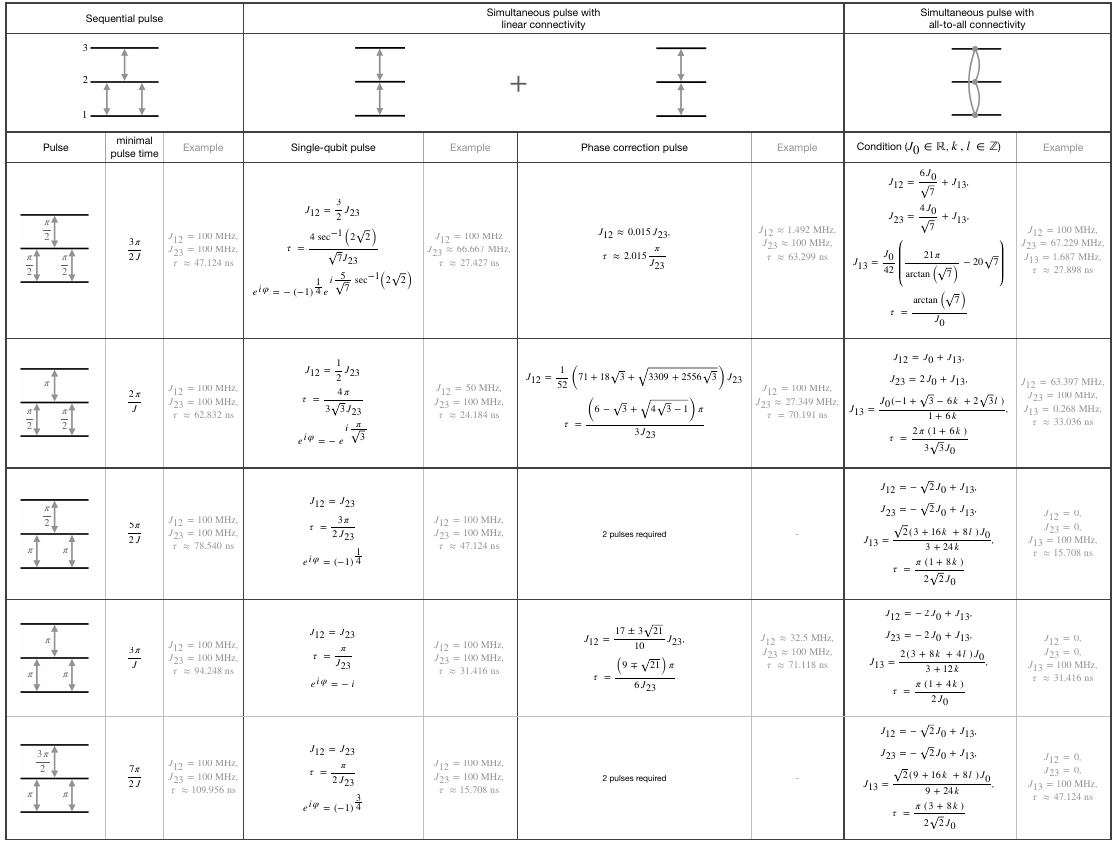}
	\caption{Sequences acting on three spins, which are frequently used in two-qubit gates \cite{FongWandzura,2Qpaper2022}. When using sequential pulses three steps are required. With linear connectivity two to three simultaneous pulses are required, where the second (and third) are purely phase corrections relative to leakage states as described in Eqs.~\eqref{Eq:phase-correction-linearQAS0} and \eqref{Eq:phase-correction-linearQAS1}. With all-to-all connectivity only one pulse is needed to generate the same sequence. For each sequence we give conditions for the exchange values $J_{12}$, $J_{23}$, ($J_{13}$) and the driving time $\tau$. For the linear case we also show the phase difference $e^{i\phi}$ between computational and leakage states. The effective time for a pulse is the time given by $\tau_{\rm eff} = \tau + \tau_{\rm idle}$, where $\tau$ is given in the table and $\tau_{\rm idle}$ is the idle time between two pulses}
	\label{Fig:2Qsubsequences}
\end{table*}

\section*{Acknowledgments}
This work has been supported by QLSI with funding from the European Union's Horizon 2020 research and innovation programme under grant agreement No.~951852.

\section*{Author contributions}
The theoretical analysis was performed by I.H. R.K. helped writing the simulation code. F.A.M provided pulse sequence timing information and the energy scales. F.B., M.C., and M.T.M. developed the measurement routine and performed the experiment. F.L. built the measurement software framework. N.B. coordinated the project and edited the draft and figures. G.B. supervised the project and helped writing the manuscript.

\appendix

\section{Charge noise sweet spots in triple quantum dots} \label{App:SweetSpot}
For triple quantum dots a charge noise sweet spot was shown for the linear connectivity \cite{PhysRevB.93.121410, PhysRevB.94.165411}. Analogously, a sweet spot for the all-to-all connectivity is obtained. Using the Hubbard model,
\begin{align}
    H=& \sum_{i=1}^3 \sum_{\sigma \in \{\downarrow, \uparrow\}} \epsilon_i \hat{c}_{i,\sigma}^{\dagger} \hat{c}_{i,\sigma} + \sum_{i=1}^{3} U_i \hat{n}_{i,\uparrow} \hat{n}_{i,\downarrow} + \sum_{i \neq j} V_{ij} \hat{n}_i \hat{n}_j \notag \\
    & +\sum_{\langle i,j \rangle} \sum_{\sigma} t_{ij} \left( \hat{c}_{i,\sigma}^{\dagger} \hat{c}_{j,\sigma} + \rm{h.c.} \right),
\end{align}
where $\hat{n}_{i} = \hat{n}_{i, \downarrow}+\hat{n}_{i,\uparrow}$ and $\hat{n}_{i,\sigma} = \hat{c}_{i,\sigma}^{\dagger} \hat{c}_{i,\sigma}$,
we obtain the exchange interactions between neighboring electron spins
\begin{align}
    J_{12} = \frac{t_{12}^2 (U_1 + U_2 + \Tilde{V}_{13} + \Tilde{V}_{23})}{(U_1 + \Tilde{V}_{13} + \epsilon_1 - \epsilon_2) (U_2 + \Tilde{V}_{23} - \epsilon_1 + \epsilon_2)}, \\
    J_{23} = \frac{t_{23}^2 (U_2 + U_3 + \Tilde{V}_{12} + \Tilde{V}_{13})}{(U_2 + \Tilde{V}_{12} + \epsilon_2 - \epsilon_3) (U_3 + \Tilde{V}_{13} - \epsilon_2 + \epsilon_3)}, \\
    J_{13} = \frac{t_{13}^2 (U_1 + U_3 + \Tilde{V}_{12} + \Tilde{V}_{23})}{(U_1 + \Tilde{V}_{12} + \epsilon_1 - \epsilon_3) (U_3 + \Tilde{V}_{23} - \epsilon_1 + \epsilon_3)}.
\end{align}
Here we defined $\Tilde{V}_{12} = V_{12} - V_{23} - V_{13}$, $\Tilde{V}_{23} = -V_{12} + V_{23} - V_{13}$ and $\Tilde{V}_{13} = -V_{12} - V_{23} + V_{13}$. We find a first order sweet spot for fluctuations $\delta \epsilon_i$ in $\epsilon_i$ if
\begin{align}
    \epsilon_2 - \epsilon_1 = \frac{U_1 - U_2 + \Tilde{V}_{13} - \Tilde{V}_{23}}{2}, \\
    \epsilon_3 - \epsilon_1 = \frac{U_1 - U_3 + \Tilde{V}_{12} - \Tilde{V}_{23}}{2}.
\end{align}

\section{Simultaneous exchange interaction in two-qubit gates} \label{App:2Qgates}

\begin{figure*}[ht]
	\centering
	\includegraphics[width=0.80\textwidth]{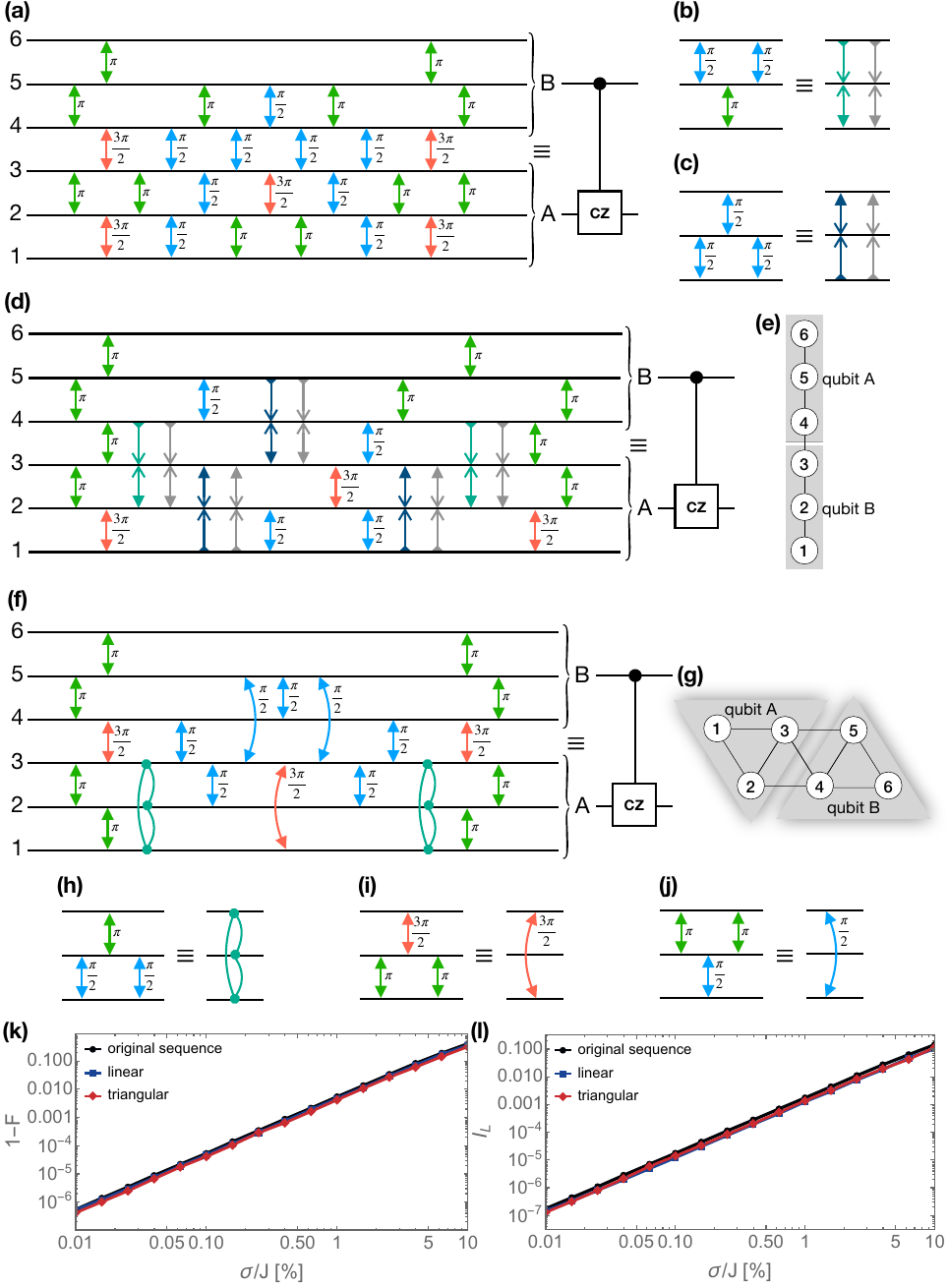}
	\caption{(a) CZ gate sequence for a linear connectivity as described in the main text, taken from Ref.~\cite{2Qpaper2022}, first introduced in \cite{FongWandzura}. Here, $B$ is the control and $A$ the target qubit. (b)-(c)~Three-step exchange sequences acting on three spins that can be replaced by two simultaneous exchange pulses. (d) CZ gate sequence equivalent to (a) using the resplacements in (b)-(c). (e)~Linear connectivity of the spins used in sequences in (a) and (d). (f)~CZ gate for the triangular connectivity in (g) obtained via transformation of the sequence in (a) by replacing the sequences in (h)-(j). (g)~Triangular connectivity: Qubit $A$ and $B$ are represented as the left and right triangular, respectively. (h)-(j)~Three-step sequences acting on three qubits that can be replaced by a single (simultaneous) pulse. (k)~Infidelity of the original sequence in (a), the alternative in linear arrangement in (d) and the triangular arrangement in (f). For the Monte Carlo simulation quasi-static Gaussian-distributed charge noise was introduce with standard deviations up to $\sigma_{ij} = 0.1 J_{ij}$. (j)~Mixing infidelity as defined in Eq.~\eqref{Eq:LeakageDef}}
	\label{Fig:CZgate}
\end{figure*}

For the Fong-Wandzura CZ gate as in Ref.~\cite{2Qpaper2022} (see Fig.~\ref{Fig:CZgate}(a)) we explicitly show how subsequences in Tab.~\ref{Fig:2Qsubsequences} can be used to replace three-step sequences. We show the original CZ gate sequence in Fig.~\ref{Fig:CZgate}(a). If we assume the same linear connectivity as in the initial sequence, shown in Fig.~\ref{Fig:CZgate}(e), we can replace the subsequences as shown in Figs.~\ref{Fig:CZgate}(b) and (c) and obtain Fig.~\ref{Fig:CZgate}(d). On the other hand, if we allow for a triangular connectivity \cite{acuna2024coherent} as shown in Fig.~\ref{Fig:CZgate}(g), we can replace the subsequences shown in Figs.~\ref{Fig:CZgate}(h)-(j) in the original sequence and obtain the CZ gate sequence in Fig.~\ref{Fig:CZgate}(f). All three sequences Figs.~\ref{Fig:CZgate}(a), (d) and (f) result in the same gate operation. We again note that the sequence was found and optimized for the brick-structured sequential sequence, The original sequence can be executed in 13 time steps, where within each time step the applied exchange interactions commute. For $J_{\rm max} = 100$~MHz the gate time is then $424.115$~ns + $13\tau_{\rm idle}$. The alternative implementation in Fig.~\ref{Fig:CZgate}(d) with linear connectivity requires 16 steps and $680.839$~ns + $16\tau_{\rm idle}$, respectively. The sequence in Fig.~\ref{Fig:CZgate}(f) requires 13 steps and 364.523~ns + $13\tau_{\rm idle}$. When counting each single pulse separately, we find (a) 26 pulses in 753.982~ns + $26\tau_{\rm idle}$, (d) 26 steps in 947.875~ns + $26\tau_{\rm idle}$ of which the phase correction takes 330.279~ns, and (f) 20 steps in 568.727~ns + $20\tau_{\rm idle}$.

We estimate the performance of all three CZ gate implementations by calculating the fidelity of each sequence in the presence of noise. 
The EO qubit is defined in a subspace of a larger Hilbert space. For two-qubit gates states outside of the computational space are used as mediators for the quantum operation. Leakage can arise from charge noise during two-qubit gates and reduces the gate fidelity. However, the resulting actual gate operation that acts entirely outside the computational space is not of particular interest, since leakage states do not carry the desired quantum information. We thus define the subspace averaged gate fidelity by projecting onto the relevant qubit space. The fidelity is defined as~\cite{PhysRevA.60.1888, PEDERSEN200747}
\begin{align}
    F =  \frac{\text{Tr} [MM^{\dagger}] + \left|\text{Tr}[M]\right|^2}{d(d+1)}, \label{Eq:FidelityDef}
\end{align}
with $M = PU^{\dagger}_{\rm ideal} P U_{\rm actual}P$, where $P$ is the projector on the subspace of interest with dimension $d$. If leakage has occurred, $M$ is no longer unitary and thus both terms in Eq.~\eqref{Eq:FidelityDef} give non-trivial contributions \cite{PEDERSEN200747}. 

For the fidelity estimation we assume quasi-static Gaussian noise for simplicity and perform a Monte Carlo simulation with standard deviations up to $\sigma_{ij}=0.1 J_{ij}$ and 1000 runs. The fidelity of the CZ gates are compared in Fig.~\ref{Fig:CZgate}(k). The original sequence is shown in black, the sequence of Fig.~\ref{Fig:CZgate}(d) with a linear arrangement in blue and the sequence for the triangular arrangement in Fig.~\ref{Fig:CZgate}(f) in red. All curves feature a quadratic slope that can be explained by the quasi-static noise approximation using Gaussian-distributed fluctuations. We find that all three sequences perform similarly and only find a slight improvement in the triangular arrangement. This can be explained by the rather similar length of all sequences, since both new variants were transformed from the original sequence in Fig.~\ref{Fig:CZgate}(a) that was optimized for the sequential execution of non-commuting exchange pairs.

To quantify the mixing between qubit and leakage states we utilize the Frobenius norm $\|A\| = \sqrt{\sum_{i,j} |a_{ij}|^2 }$ of a matrix $A$. For a unitary matrix $U$ we have $\|U\|^2=d$. We thus define the mixing infidelity by
\begin{align}
    I_L=1-\frac{\|P_Q U\|^2 + \|P_L U\|^2}{d_Q + d_L} , \label{Eq:LeakageDef}
\end{align}
where $P_Q$ and $P_L$ are the projections onto the qubit and leakage spaces with dimensions $d_Q$ and $d_L$, respectively. If $U$ does not introduce any leakage, we have $ \|P_Q U\|^2 = d_Q$ and $ \|P_L U\|^2 = d_L$, and thus $I_L =0$. However, if $U$ contains block-off-diagonal elements, it introduces transitions between the two spaces and $ \|P_Q U\|^2 < d_Q$ and $ \|P_L U\|^2 < d_L$, and hence $I_L >0$.
Here we neglect the distinction between leakage and seepage, which describe the change of population from qubit to leakage space and vice versa. 
The mixing infidelity for the Monte Carlo simulations is shown in Fig.~\ref{Fig:CZgate}(l). Again we find a quadratic slope of all curves and only a very slight improvement for the alternative sequences compared to the original one.

Although our results only show a slight decrease in the infidelity for a triangular connectivity, we have shown how the total number of pulses reduces from 26 to 20 which also minimizes the number of idling times between two exchange pulses. More simplifications can be made by optimizing the sequence for simultaneous exchange pulses.

\bibliography{bibliography}

\end{document}